\documentclass[
 reprint,
superscriptaddress,
nofootinbib,
 amsmath,amssymb,
 aps,
pra,
]{revtex4-2}

\usepackage{amsmath,amssymb,amsfonts,bm}
\usepackage{slashed}
\usepackage{graphicx}
\usepackage{hyperref}
\usepackage{subfigure}
\usepackage{tikz}
\usepackage{tikz-feynman}
\usepackage{mathtools}
\usepackage{bbold}

\newcommand{\e}{\mathrm{e}}

\definecolor{lime}{HTML}{A6CE39}
\DeclareRobustCommand{\orcidicon}{%
	\begin{tikzpicture}
	\draw[lime, fill=lime] (0,0) 
	circle [radius=0.16] 
	node[white] {{\fontfamily{qag}\selectfont \tiny ID}};
	\draw[white, fill=white] (-0.0625,0.095) 
	circle [radius=0.007];
	\end{tikzpicture}
	\hspace{-2mm}
}

\foreach \x in {A, ..., Z}{%
	\expandafter\xdef\csname orcid\x\endcsname{\noexpand\href{https://orcid.org/\csname orcidauthor\x\endcsname}{\noexpand\orcidicon}}
}

\begin{document}

\preprint{APS/123-QED}

\title{Unruh-induced decoherence and entanglement degradation in Ramsey interferometry}

\author{E. D. C Brand\~{a}o\orcidA{}}
\email{edson.brandao@ufpi.edu.br}
 \affiliation{Departamento de Física, Universidade Federal do Piauí, Teresina, 64049-550, Piauí, Brazil}

\author{O. P. de Sá Neto\orcidD{}}
\email{olimpiopereira@phb.uespi.br}
 \affiliation{Campus Prof. Alexandre Alves de Oliveira, Universidade Estadual do Piauí, Parnaíba, Piauí, Brazil}

\author{F. C. E. Lima\orcidC{}}
\email{cleiton.estevao@ufabc.edu.br}
\affiliation{Centro de Matemática, Computação e Cognição (CMCC), Universidade Federal do ABC (UFABC), Av. dos Estados 5001, CEP 09210-580, Santo André, São Paulo, Brazil}

\author{H. A. S. Costa\orcidB{}}%
 \email{hascosta@ufpi.edu.br (Corresponding author)}
\affiliation{Departamento de Física, Universidade Federal do Piauí, Teresina, 64049-550, Piauí, Brazil}

\begin{abstract}
We propose a modified Ramsey interferometric protocol to probe decoherence associated with the Unruh effect through fringe visibility. The scheme employs Unruh-DeWitt detectors coupled to a massless scalar field in the Minkowski vacuum through an interaction that induces pure dephasing without transitions between their internal energy levels. For a uniformly accelerated detector with finite spatial extent, an analytical dephasing function separates inertial vacuum fluctuations from the thermal Unruh contribution. When the thermal contribution dominates, fringe visibility decays exponentially with a rate proportional to the proper acceleration. The bipartite extension considers two detectors initially prepared in a Bell state, one inertial and the other uniformly accelerated, while retaining correlations mediated by the scalar field. In particular, we find that the joint fringe visibility coincides exactly with concurrence  when the cross-term contribution disappears. This suggests that the entanglement degradation can be quantified directly from the interference contrast. 
\end{abstract}

                             
\maketitle
\newpage

\section{Introduction}

The connection between Quantum Field Theory (QFT) and Relativity raises fundamental questions about how different observers describe the same quantum system \cite{fulling1973nonuniqueness,davies1975scalar,unruh1976notes,barros2025mitigating}. One striking consequence is that the notion of a particle depends on the observer's state of motion \cite{Bire82}. This feature is particularly evident in the Unruh effect \cite{unruh1976notes}, which predicts that an observer moving with constant proper acceleration $a$ experiences the Minkowski vacuum as a thermal state with temperature $T_U=\hbar a/(2\pi k_B c)$ \cite{crispino2008unruh,takagi1986vacuum,paredes2009temperature}. This effect reflects a fundamental distinction between inertial and accelerated descriptions of quantum fields and is closely related to the causal horizon along uniformly accelerated trajectories \cite{rindler1966kruskal}. 

Although the Unruh effect is firmly established within quantum field theory, obtaining direct experimental confirmation remains a significant challenge \cite{lochan2020detecting,lynch2021experimental,gregori2024measuring}. The main challenge is the extremely low temperature associated with accelerations attainable under laboratory conditions \cite{wang2014quantum,cozzella2017proposal,stargen2022cavity,toussaint2026detecting}. For instance, an Unruh temperature of approximately $1\,\mathrm{K}$ would require an acceleration of order $10^{20}\,\mathrm{m}/\mathrm{s}^2$ \cite{wang2014quantum,cozzella2017proposal,stargen2022cavity,toussaint2026detecting,parry2025waiting}. Such demanding conditions have encouraged the search for alternative signatures of the effect. Therefore, instead of focusing solely on directly detecting a thermal spectrum, one may explore how acceleration influences the dynamics of a quantum system interacting with vacuum fluctuations. This perspective has attracted considerable interest in relativistic quantum information, where localized quantum detectors provide a natural framework for studying how accelerated motion affects coherence and entanglement \cite{liu2022does,liu2023entanglement,perche2024particle,wu2025entangled}. A standard framework is the Unruh-DeWitt model \cite{unruh1981experimental,dewitt1979quantum}, in which a two-level system interacts with a quantum field along a prescribed trajectory. Most conventional treatments focus on detector excitation and de-excitation, since transition probabilities encode information about the field fluctuations experienced by the moving observer \cite{louko2008transition,pozas2015harvesting}. However, these fluctuations can also affect the relative phase between quantum states, even when transitions between the energy levels are absent. This insight opens a novel approach to probing acceleration-induced effects by focusing on the decay of quantum coherence.

In order to explore this possibility, we consider a detector–field interaction that produces pure dephasing. The interaction Hamiltonian commutes with the detector's free Hamiltonian, leaving the energy-level populations unchanged while suppressing the off-diagonal elements of its reduced density matrix. The resulting dynamics separate the dephasing of quantum coherence from population relaxation. They also enable distinguishing the contribution already present for an inertial detector from the additional decoherence associated with acceleration. Ramsey interferometry is particularly well suited to this setting, since the coherence of a two-level system is directly reflected in the visibility of its interference fringes \cite{costa2020ramsey, lopes2021thermal, barros2023robustness}. 

An important outcome of the two-detector analysis is the relationship between interferometric visibility and quantum entanglement, a similar result has been reported in \cite{lee2008proposal}. For the family of states considered, the pure-dephasing interaction preserves the initial Bell state, while its coherence decreases through coupling to the field. In particular, at the spatial decoupling limit of the correlations between the detectors, when the cross-term contribution disappears. We find an analytical relationship between visibility and concurrence. Therefore, the joint visibility serves as an immediate quantitative indicator of the residual bipartite entanglement. This result connects the degradation of quantum correlations under accelerated motion to an observable quantity obtained from an interferometric measurement. 

The paper is organized as follows. Sect. II presents the detector–field model and derives the pure-dephasing dynamics for a uniformly accelerated trajectory. Sect. III discusses the single-detector Ramsey protocol and the behavior of its interference visibility. Furthermore, we also extend the treatment to two entangled detectors, examine the role of field-mediated correlations, and establish the connection between joint visibility and concurrence. Sect. IV announces our findings. Throughout this work, the metric signature (+ - - -) and natural units ($\hbar=c=k_B=1$) are used.

\section{Phase decoherence of a single qubit}

 Let us begin this section by briefly reviewing the quantization of the scalar field in Rindler spacetime. It is well appreciated that a uniformly accelerated observer in the $(t, z)$ plane has a hyperbolic world line in Minkowski spacetime. In Rindler coordinates $(\tau, \zeta)$ this word line is described by two different sets coordinates:
 \begin{align}
 \text{Region I}: \nonumber \\
     t(\tau) &= \frac{\e^{a\zeta}}{a}\sinh(a\tau), \quad  z(\tau) = \frac{\e^{a\zeta}}{a}\cosh(a\tau), \label{Rindler1} \\
 \text{Region II:} \nonumber \\
     t(\tau) &= -\frac{\e^{a\zeta}}{a}\sinh(a\tau), \quad  z(\tau) = -\frac{\e^{a\zeta}}{a}\cosh(a\tau), \label{Rindler2}
 \end{align}
 where Region I and Region II are the portion $z > |t|$ (right Rindler wedge) and $z < |t|$ (left Rindler wedge) of the Minkowski spacetime, repectively. $a$ denotes detector's proper acceleration. By using (\ref{Rindler1}) or (\ref{Rindler2}), the line element restricted to the Rindler wedge is given by
 \begin{align*}
     ds^2 = \e^{2a\zeta}(d\tau^2-d\zeta^2)-d\mathbf{x}_\perp^2,
 \end{align*}
 where $\mathbf{x}_\perp=(x,y)$ are the coordinates transverse to the direction of acceleration. The quantization scheme of scalar field $\phi(x)$ in the Rindler wedge follows these steps \cite{pringle1989rindler}: First, consider that $\phi(x)$ satisfying the massless Klein-Gordon equation
 \begin{align}
     \square \phi = 0,
 \end{align}
 whose a complete set of solutions in the Rindler wedge is given by
 \begin{align}
v_{\omega \textbf{k}_\perp}(\tau, \textbf{r}) &= \sqrt{\frac{4\omega \sinh(\pi\omega/a)}{\pi a (2\pi)^2}} \nonumber \\
&\times K_{i\omega/a}\left(\frac{k_\perp}{a} \e^{a\zeta}\right) \e^{i \mathbf{k}_\perp \cdot \mathbf{x}_\perp-i\omega\tau}, \label{vk}
\end{align}
 where $\textbf{r} = (\zeta, \textbf{x}_\perp)$ and $K_{i\omega/a}$ is the modified Bessel function of the second kind. The modes defined in Eq. (\ref{vk}) are said to be positive-frequency with respect to timelike Killing vector $\partial_{\tau}$ since $\partial_\tau v_{\omega k_\perp} = -i\omega v_{\omega k_\perp}$ with $\omega > 0$. Besides, the modes function $v_{\omega \textbf{k}_\perp}$ are orthogonal and normalized with respect to the Klein-Gordon inner product,
 \begin{align}
     (v_{\omega \textbf{k}_\perp}, v_{\omega' \textbf{k}'_\perp}) = \delta(\omega - \omega')\delta(\textbf{k}_\perp - \textbf{k}'_\perp).
 \end{align}
 Then, in terms of the Rindler modes Eq. (\ref{vk}), the scalar field can be expressed as
\begin{align} \label{eq:campo_modos}
    \phi(t, \mathbf{x}) = \sum_{\omega\textbf{k}_\perp}\frac{1}{\sqrt{2\omega V}} \left[ v_{\omega \textbf{k}_\perp}b_{\omega \textbf{k}_\perp} + v_{\omega \textbf{k}_\perp}^* b_{\omega \textbf{k}_\perp}^\dagger\right],
\end{align}
where $b_{\omega \textbf{k}_\perp}$ and $b_{\omega \textbf{k}_\perp}^\dagger$ are the annihilation and creation operators, respectively, of Rindler modes. These operators obey the usual canonical commutation relations $[b_{\omega\textbf{k}_\perp},b_{\omega'\textbf{k}'_\perp}^\dagger]=\delta_{\omega\omega'}\delta_{\textbf{k}_\perp\textbf{k}'_\perp}$ and all the other commutators vanish.
 
 
  Now let us analyze the behavior of a uniformly accelerated detector in the Minkowski spacetime. Recently, Nesterov and collaborators proposed an alternative detector model to investigate the Unruh effect via the decoherence of the reduced density matrix \cite{nesterov2020decoherence}. According to the authors, assuming that the detector is located at the origin of the coordinates, the Hamiltonian that describes the interaction between a two-level detector with a frequency gap $\Omega$ and a massless scalar field $\phi(x)$ can be written as
 \begin{align}
     H = H_{\phi} + H_d + H_{\mathrm{int}},
 \end{align}
 where $H_{\phi} $ is the Klein-Gordon Hamiltonian for the massless scalar field. $H_d $ is the detector's Hamiltonian, given by
 \begin{align*}
     H_d = \frac{\Omega}{2}\sigma_z.
 \end{align*}
 Here $\sigma_z=\vert e\rangle\langle e\vert-\vert g\rangle\langle g\vert$, and $\vert e\rangle$ and $\vert g\rangle$ are the excited and ground states of the detector, respectively. The interaction Hamiltonian can be written as 
\begin{align}
\label{eq:hint_geral}
    H_{\text{int}} &= \lambda\int_{\sum_\tau} d^3\textbf{r}\sqrt{-g}f(\textbf{r})\sigma_z \otimes \phi(t, \mathbf{x}), \nonumber \\
    &= \lambda\sigma_z\otimes \sum_{\omega\textbf{k}_\perp}\frac{1}{\sqrt{2\omega V}}\left[ g_{\omega \textbf{k}_\perp}^* b_{\omega \textbf{k}_\perp} + g_{\omega \textbf{k}_\perp} b_{\omega \textbf{k}_\perp}^\dagger\right],
\end{align}
where $g \equiv \text{det}(g_{\mu\nu})$ with $g_{\mu\nu}$ being the metric tensor of Rindler spacetime, $\lambda$ is a coupling constant, $\sum_\tau$ represent the Cauchy surface associated with some timelike isometry, the function $f(\textbf{r})$ models the spatial profile of the detector in the Rindler coordinates, and $g_{\omega \textbf{k}_\perp}$ is a compact complex function defined as
\begin{align}
    g_{\omega \textbf{k}_\perp} = \int_{\sum_\tau} d^3\textbf{r}\sqrt{-g}f(\textbf{r})v_{\omega \textbf{k}_\perp}.
\end{align}
In the interaction picture, the state that describes the detector and the scalar field at proper time $\tau$ is given by
\begin{align}
    |\psi(\tau)\rangle = \textbf{T}\exp\left[-i\int_{0}^{\tau}d\tau' H_{\text{int}}^{I}(\tau')\right]|\psi(\tau_0)\rangle 
\end{align}
where $\textbf{T}$ is the time-ordering operator and $H_{\text{int}}^{I}(t) = U_0^{\dagger}(\tau)H_{\text{int}}(\tau)U_0(\tau)$ with $U_0(\tau)$ being the unitary evolution operator associated with $H_{\phi} + H_d$. The final state of the system (detector plus field) can be expressed as
  \begin{align*}
      |\psi(\tau)\rangle &= \textbf{T}\exp\Bigg[-i\nu(\tau)\mathbb{1} + \frac{i\Omega}{2}\sigma_z\otimes \mathbb{1} \nonumber \\
      &+ \sigma_z\otimes\sum_{\omega\textbf{k}_\perp}\left[ \xi_{\omega \textbf{k}_\perp}(\tau) b_{\omega \textbf{k}_\perp}^{\dagger} - \xi_{\omega \textbf{k}_\perp}^*(\tau) b_{\omega \textbf{k}_\perp}\right] \Bigg]|\psi(\tau_0)\rangle,
  \end{align*}
where $\nu(\tau) = \frac{i}{2}\sum_{\omega\textbf{k}_\perp}\int_{0}^{\tau}d\tau'[\xi_{\omega \textbf{k}_\perp}^*(\tau')\dot{\xi}_{\omega \textbf{k}_\perp}(\tau') - \xi_{\omega \textbf{k}_\perp}(\tau')\dot{\xi}_{\omega \textbf{k}_\perp}^*(\tau')]$ with $\xi_{\omega \textbf{k}_\perp}(\tau) = \lambda g_{\omega \textbf{k}_\perp}\frac{1 - e^{-i\omega\tau}}{\sqrt{2V}\omega^{3/2}}$. In particular, consider that the system is initially prepared in the state 
\begin{align*}
   \vert\psi(0)\rangle=\frac{1}{\sqrt{2}}\left(\vert e\rangle+\vert g\rangle\right)\otimes\vert 0_M\rangle, 
\end{align*}
 where $|0_M\rangle$ correspond to the Minkowski vacuum of the scalar field. Since the quantum fluctuations of the field are conditionally excited according to the state of the detector, at the end of the interaction time, the global state evolves into a coherent superposition of field states
\begin{align*}
|\psi(\tau)\rangle &= \frac{1}{\sqrt{2}}\bigg[\e^{-i\nu(\tau) - i\Omega\tau/2} |e\rangle \otimes\vert\Psi_e(\tau)\rangle \\
&+ \e^{-i\nu(\tau) + i\Omega\tau/2} |g\rangle \otimes |\Psi_g(\tau)\rangle \bigg],
\end{align*}
where $\vert\Psi_e(\tau)\rangle = \prod_{\omega \textbf{k}_\perp} D_{\omega \textbf{k}_\perp}(\xi_{\omega \textbf{k}_\perp})|0_M\rangle$ and $|\Psi_g(\tau)\rangle = \prod_{\omega \textbf{k}_\perp} D_{\omega \textbf{k}_\perp}(-\xi_{\omega \textbf{k}_\perp})|0_M\rangle$ are the coherent field states induced by the interaction, with $D_k(\xi)=\e^{\xi_{\omega \textbf{k}_\perp} b_{\omega \textbf{k}_\perp}^\dagger - \xi_{\omega \textbf{k}_\perp}^*b_{\omega \textbf{k}_\perp}}$. By taking the partial trace over the field Fock degrees of freedom, the reduced density matrix of the detector is given by
\begin{align*}
   \rho(\tau) &= \text{Tr}_{\phi}\left[|\psi(\tau)\rangle\langle\psi(\tau)|\right], \\
   &= \frac{1}{2}\begin{pmatrix}
       \rho_{ee}(\tau) & \rho_{eg}(\tau)  \\
      \rho_{ge}(\tau)   & \rho_{gg}(\tau) 
   \end{pmatrix},
\end{align*}
with
\begin{align*}
    \rho_{ee}(\tau) &= \rho_{gg}(\tau) = 1, \\
    \rho_{eg}(\tau)  &= \rho_{ge}^*(\tau) = \e^{-i\Omega\tau - \gamma(\tau)},
\end{align*}
where the decoherence function $\gamma(\tau) = 2 \sum_k |\xi_k(\tau)|^2 \coth\left(\frac{\pi\omega_k}{a}\right)$, see Ref. \cite{nesterov2020decoherence} for more details. In the continuum limit, the decoherence function is defined by the integral
\begin{align}
\label{eq:gamma_integral_espectral}
\gamma(\tau) = \frac{2\lambda^2}{\pi^3} \int_0^\infty \frac{|g_{\omega}|^2}{\omega}\sin^2\left(\frac{\omega\tau}{2} \right) \coth\left(\frac{\pi\omega}{a}\right) d\omega.
\end{align}
 Here $|g_{\omega}|^2 = \int d^2\textbf{k}_\perp |g_{\omega \textbf{k}_\perp}|^2$. The regularized form of the decoherence function can be written as 
\begin{align*} 
\gamma(\tau) = \frac{4\lambda^2}{\pi^2} \int_0^\infty \frac{\e^{-l\omega}}{\omega} \sin^2\left(\frac{\omega\tau}{2}\right) \coth\left(\frac{\pi\omega}{a}\right) d\omega.
\end{align*}
where $l$ represents the size of the detector. When $l$ goes to zero, it becomes a point-like detector. Performing the integration, we obtain
\begin{align}
\gamma(\tau) = \gamma_{\text{vac}}(\tau) + \gamma_{\text{th}}(\tau),
\end{align}
where
\begin{align*}
    \gamma_{\text{vac}}(\tau) &= \frac{2\lambda^2}{\pi^2}\,\operatorname{Re}\,\operatorname{ln}\left(1+i\frac{\tau}{l}\right) \\
    \gamma_{\text{th}}(\tau) &= -\frac{4\lambda^2}{\pi^2} \operatorname{Re} \ln \left[ \frac{\Gamma\left(1 + \frac{al}{2\pi} + i\frac{a\tau}{2\pi}\right)}{\Gamma\left(1 + \frac{al}{2\pi}\right)} \right].
\end{align*}
Here $\Gamma(z)$ is the Gamma function. In the asymptotic regime od large proper times ($\tau \gg l$) and high accelerations  ($a\tau\gg 1$), we can use Stirling’s approximation for the Gamma function's behavior in the complex plane. This allow us to approximate the decoherence function as 
\begin{align}
    \gamma(\tau) \approx \frac{2\lambda^2}{\pi^2} \ln\left(\frac{\tau}{l}\right) + \frac{\lambda^2 a \tau}{\pi^2}.
\end{align}
This result reveals the coexistence of two additive and independent physical mechanisms of phase decay. The first, logarithmic term represents the minimal inertial dephasing generated by Minkowski zero-point quantum fluctuations. The second term, which is linear in both the proper time $\tau$ and the acceleration $a$, represents the noninertial decoherence of thermal origin induced by the Unruh effect, and dominates the dephasing process in the regime of high accelerations \cite{nesterov2020decoherence}.


\section{Ramsey interferometric circuit}

\subsection*{Setup A}

\begin{figure}[!h]
    \centering
    \includegraphics[width=0.45\textwidth]{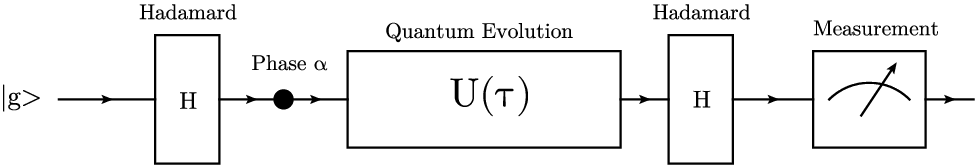}\vspace{-0.2cm}
    \vspace{0.5cm}
    \caption{Ramsey interferometric circuit for a single detector that is uniformly accelerated between two Hadamard gates.}
    \label{fig:esquema_ramsey_1_qubit}
\end{figure}

In this section, we analyze the loss of quantum coherence induced by the Unruh effect in a Ramsey interferometric circuit, as illustrated in Fig. \ref{fig:esquema_ramsey_1_qubit}. It well known that a Ramsey interferometer probes the relative phase and coherence of a two-level system through two $\pi/2$ pulses separated by a free-evolution interval; the resulting fringe visibility provides a direct measure of the off-diagonal coherence of the reduced density matrix \cite{ilikj2025ramsey}. 

In this framework, the circuit operates in the composite Hilbert space $\mathcal{H} = \mathcal{H}_d \otimes \mathcal{H}_{\phi} $, with the detector initially prepared in its ground state and the real massless scalar field in the inertial Minkowski vacuum, $\vert\Psi_{\mathrm{in}}\rangle=|g\rangle\otimes|0_M\rangle$. The first Hadamard gate, prepares the detector in the coherent superposition 
\begin{align*}
  \vert \psi_1\rangle = \frac{1}{\sqrt{2}}\left(|g\rangle+|e\rangle\right)\otimes|0_M\rangle. 
\end{align*}
 Then, a controllable classical phase shift $\alpha$ is applied to the excited-state branch, while the detector follows a hyperbolic trajectory with constant proper acceleration $a$ over the proper-time interval $\tau\in[0,T]$. Evolution under the conservative dephasing Hamiltonian correlates the qubit dynamics with displaced coherent states of the quantum field, so that the global state evolves into 
 \begin{align*}
     \vert\psi_2\rangle = \frac{1} {\sqrt{2}}
\left[|g\rangle\otimes|\Psi_g(\tau)\rangle+\e^{-i\theta}|e\rangle\otimes|\Psi_e(\tau)\rangle\right],
 \end{align*}
 where $\theta = \Omega\tau+\alpha$ is the total accumulated phase. The final state of the system after the operation of the second Hadamard gate is given by
  \begin{align}
     \vert\psi_3\rangle &= \e^{-\frac{i\theta}{2}}
\bigg[\cos\left(\frac{\theta}{2}\right)|g\rangle\otimes|\Psi_g(\tau)\rangle \nonumber \\
&+ i\sin\left(\frac{\theta}{2}\right)|e\rangle\otimes|\Psi_e(\tau)\rangle\bigg]. \label{psi3}
 \end{align}
 From Eq. (\ref{psi3}) we find that the probability for recording the detector in the ground state $|g\rangle$ is given by
 \begin{align}
     P_g(\tau) = \frac{1}{2}\left[1 + V(\tau)\cos(\Omega\tau+\alpha)\right].
 \end{align}
 The probability $P_g(\tau)$ exhibits one-particle interference fringes. The visibility of this interference pattern is given by $V(\tau)\equiv \e^{-\gamma(\tau)}$, which relates the Ramsey-fringe contrast to the loss of quantum phase coherence. 
 
\begin{figure}[!h]
    \centering
    \includegraphics[width=0.45\textwidth]{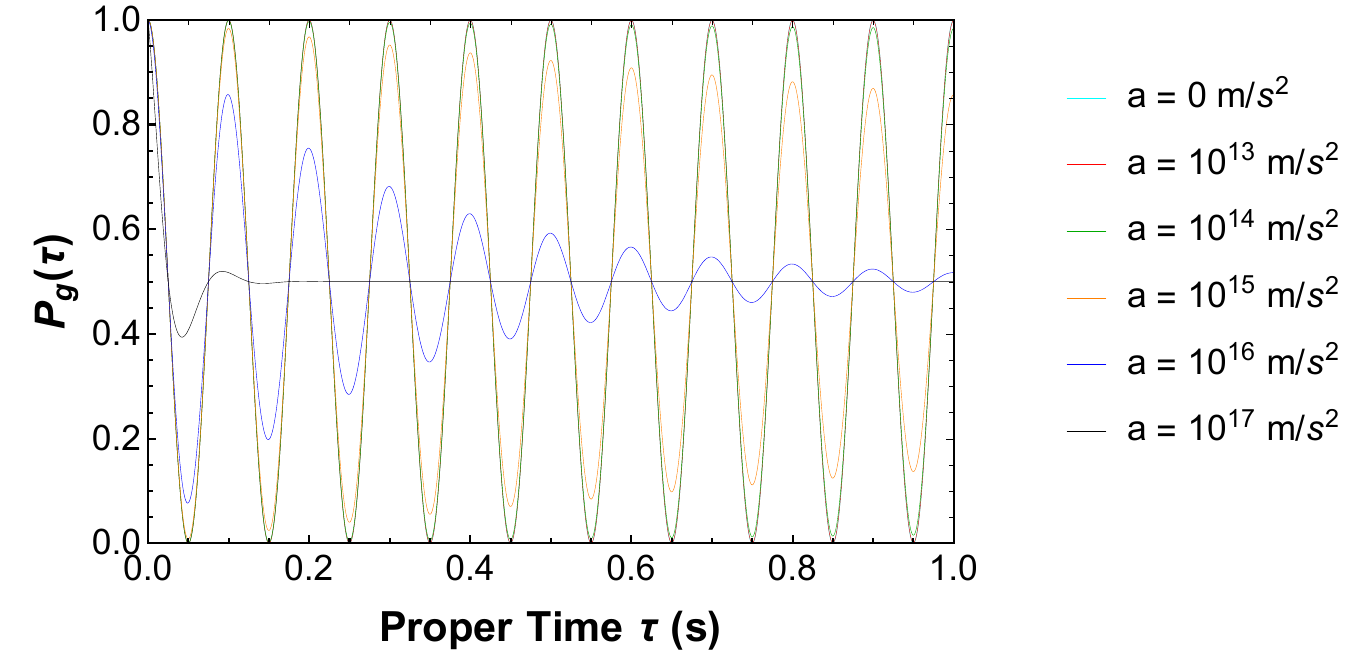}
    \caption{Transition probability $P_g(\tau)$ as a function of the proper time for different values of the  acceleration parameter $a$. Here, we adopt the coupling constant $\lambda=10^{-3}$, characteristic detector size $l \approx r_0 = 0.053 \text{ nm}$ (Bohr radius) in natural units ($c=1$), and fixed phase angle $\alpha=0$.}
    \label{prob1}
\end{figure}

 In Fig. \ref{prob1} we plot the transition probability $P_g(\tau)$ as a function of proper time for different values of the acceleration parameter $a$. In the inertial case ($a=0$), the probability displays sustained oscillations with maximal amplitude, indicating preserved phase coherence. Increasing the acceleration progressively reduces the oscillation amplitude because of the detector's interaction with Rindler excitations. For $a\geq 10^{15}\,\mathrm{m/s^2}$, the oscillations are strongly suppressed, and $P_g(\tau)$ rapidly approaches the stationary value $0.5$, signaling a substantial loss of coherence.

\begin{figure}[!h]
    \centering
    \includegraphics[width=0.45\textwidth]{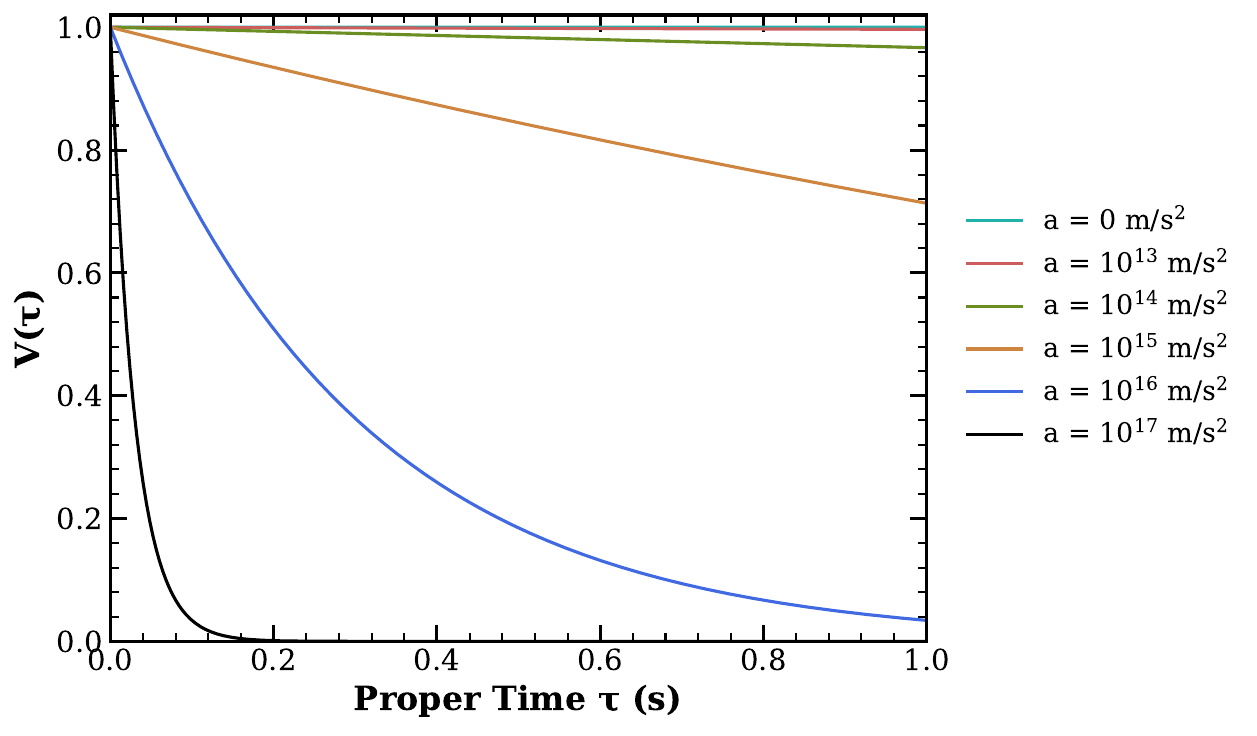}
    \caption{Visibility as a function of the proper time $\tau$ for different values of acceleration parameter $a$. Here, we adopt the coupling constant $\lambda = 10^{-3}$ and characteristic detector size $l \approx r_0 = 0.053 \text{ nm}$ (Bohr radius) in natural units ($c=1$).}
    \label{fig:visibilidade1}
\end{figure}

In Fig. \ref{fig:visibilidade1} we plot the visibility as a function of the proper-time interval $\tau\in[0.0001\,\mathrm{s},\;1.0\,\mathrm{s}]$ for different values of acceleration parameter $a$. In the absence of acceleration ($a=0$), phase degradation arises solely from Minkowski vacuum fluctuations and is governed by the logarithmic contribution $\gamma_{\mathrm{vac}}(\tau)=\frac{\lambda^2}{\pi^2}\ln(1+\tau^2/r_0^2)$. Because the coupling is weak ($\lambda=10^{-3}$), the resulting inertial dephasing remains negligible, and the visibility stays close to unity throughout the considered time interval. For nonzero acceleration ($a>0$), the Unruh thermal spectrum introduces an additional acceleration-dependent contribution to the dephasing function, $\gamma_{\mathrm{th}}(\tau)$. 

As shown in Fig. \ref{fig:visibilidade1}, when $a\leq10^{14}\,\mathrm{m/s^2}$ the visibility decreases only weakly over the $1\,\mathrm{s}$ time scale. Conversely, for $a\geq10^{15}\,\mathrm{m/s^2}$, the acceleration rapidly suppresses the interferometric contrast exponentially, driving the visibility toward zero.  These results suggest that phase decoherence induced by the Unruh thermal bath can be inferred from visibility.


\subsection*{Setup B}

\begin{figure}[!ht] 
    \centering 
    \includegraphics[width=0.45\textwidth]{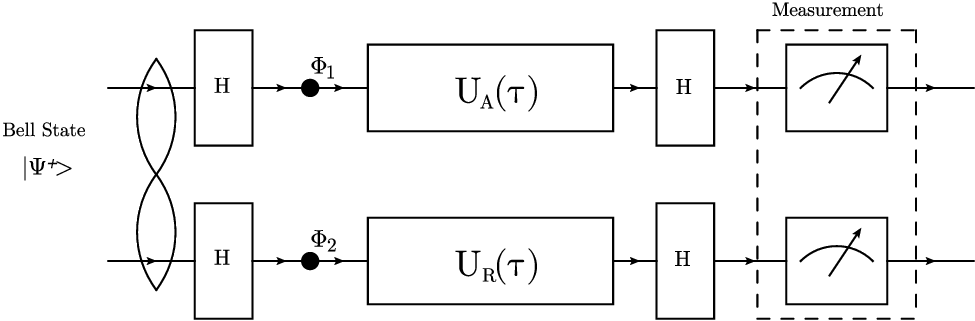}
    \vspace{0.5cm}
    \caption{Bipartite Ramsey interferometric circuit. The Alice's detector $A$ and the Rob's detector $R$  are prepared in an initially entangled bipartite state. Then, each detector passes through a local Ramsey interferometric circuit. Followed by final recombination and coincidence detection. In this setup, the Rob's detector only is uniformly accelerated between two Hadamard gates. } 
    \label{fig:esquema_ramsey_2_qubits} 
\end{figure}

In order to investigate how acceleration modifies quantum entanglement and joint visibility, we extend the interferometric protocol to a bipartite setup involving two Unruh–DeWitt detectors, Alice ($A$) and Rob ($R$). The composite system is described by the Hilbert space $\mathcal{H}=\mathcal{H}_A\otimes\mathcal{H}_R\otimes\mathcal{H}_{\phi}$. Each detector is coupled to its own local Ramsey interferometer. Detector $A$ follows an inertial trajectory in Minkowski spacetime, whereas detector $R$ undergoes constant proper acceleration $a$ along the $z$ direction between two Hadamard gates. The bipartite circuit architecture is illustrated in Fig. \ref{fig:esquema_ramsey_2_qubits}.

 First, consider two qubits prepared in the maximally entangled Bell state, $|\Psi^+\rangle$, and a massless scalar field in the Minkowski vacuum, $|0_M\rangle$. Thus, the initial state of the system is given by
\begin{align*}
|\psi_1^{\text{AR}\phi}\rangle = \frac{1}{\sqrt{2}}\left[\vert e_A g_R\rangle + \vert g_A e_R\rangle\right]\otimes |0_M\rangle.
\end{align*}
 The action of the ﬁrst Hadamard gate transforms this state into:
 \begin{align*}
|\psi_2^{\text{AR}\phi}\rangle = \frac{1}{\sqrt{2}}\left[\vert g_A g_R\rangle - \vert e_A e_R\rangle\right]\otimes |0_M\rangle.
\end{align*} 
 Then, a relative phase shift between the ground state ($\vert g_A\rangle$ or $\vert g_R\rangle$) and excited state ($\vert e_A\rangle$ or $\vert e_R\rangle$) into each interferometers is introduced to get the state
 \begin{align*}
|\psi_3^{\text{AR}\phi}\rangle = \frac{1}{\sqrt{2}}\left[\vert g_A g_R\rangle - e^{-i\Delta\alpha}\vert e_A e_R\rangle\right]\otimes |0_M\rangle.
\end{align*}
where $\Delta\alpha = \alpha_1 + \alpha_2$. Alice's detector remains inertial while Rob's detector is linearly accelerated. Each detectors couple to the massles scalar field through a conservative pure-dephasing interaction, with the local interaction Hamiltonian given by 
\begin{align}
  H_{\text{int}}^{(j)} &= \lambda\int_{\sum_\tau} d^3\textbf{r}\sqrt{-g}f(\textbf{r})\sigma_z^{(j)} \otimes \phi^{(j)}(t, \mathbf{x}), \label{HintAR}
\end{align}
with $j \in \{A, R\}$. Since $[H_{\mathrm{int}}^{(j)}, \sigma_z^{(j)}] = 0$, there is no net energy exchange with the field, so that the diagonal populations remain unchanged, while the interaction acts to damp the off-diagonal coherences. After the interaction, the system evolves into the state
\begin{align*}
|\psi_4^{\text{AR}\phi}\rangle = \frac{1}{\sqrt{2}} \left[| g_A g_R\rangle\otimes|\Psi_{gg}\rangle - e^{-i(\theta_{A} + \theta_{R})}| e_A e_R\rangle\otimes|\Psi_{ee}\rangle \right].
\end{align*}
where $\theta_j = \Omega_j\tau_j + \alpha_j$ with $j \in \{A, R\}$. Here 
\begin{align*}
 |\Psi_{ee}\rangle &= \prod_{\omega \textbf{k}_\perp} D_{\omega \textbf{k}_\perp}(\xi_{\omega \textbf{k}_\perp}^A + \xi_{\omega \textbf{k}_\perp}^R)|0_M\rangle, \\
 |\Psi_{gg}\rangle &= \prod_{\omega \textbf{k}_\perp} D_{\omega \textbf{k}_\perp}(-\xi_{\omega \textbf{k}_\perp}^A - \xi_{\omega \textbf{k}_\perp}^R)|0_M\rangle.
\end{align*}
Finally, after the operation of the second Hadamard gate, the state of the system is
\begin{align*}
&|\psi_5^{\text{AR}\phi}\rangle = \frac{1}{2\sqrt{2}} \\
&\times \bigg[\bigg(|g_A g_R\rangle + |g_A e_R\rangle + |e_A g_R\rangle + |e_A e_R\rangle \Bigg)\otimes|\Psi_{gg}\rangle \\ 
&- e^{-i(\theta_{A} + \theta_{R})} \\
&\times\Bigg(|g_A g_R\rangle - |g_A e_R\rangle - |e_A g_R\rangle + |e_A e_R\rangle \bigg)\otimes|\Psi_{ee}\rangle \bigg].
\end{align*}
Since we has no access to field modes, the reduced density matrix of detectors is obtained by tracing out the field degrees of freedom $\rho^{\text{AR}}(\tau) = \operatorname{Tr}_{\phi} \left[ \rho^{\text{AR}\phi}\right]$ with $\rho^{\text{AR}\phi} = |\psi_5^{\text{AR}\phi}\rangle\langle\psi_5^{\text{AR}\phi}|$. In the  basis $\{\vert e_A e_R\rangle,\vert e_A g_R\rangle,\vert g_A e_R\rangle,\vert g_A g_R\rangle\}$, we can write $\rho^{\text{AR}}(\tau)$ in matrix form as
\begin{align*} 
    \rho^{\text{AR}}(\tau) = \frac{1}{4}
    \begin{pmatrix} 
     1 - \text{Re}\mathcal{K} & i\text{Im}\mathcal{K} & i\text{Im}\mathcal{K} & 1 - \text{Re}\mathcal{K} \\ 
     -i\text{Im}\mathcal{K} & 1 + \text{Re}\mathcal{K} & 1 + \text{Re}\mathcal{K} & -i\text{Im}\mathcal{K} \\ 
     -i\text{Im}\mathcal{K} & 1 + \text{Re}\mathcal{K} & 1 + \text{Re}\mathcal{K} & -i\text{Im}\mathcal{K} \\
     1 - \text{Re}\mathcal{K} & i\text{Im}\mathcal{K} & i\text{Im}\mathcal{K} & 1 - \text{Re}\mathcal{K} 
     \end{pmatrix}, 
\end{align*}
where $\mathcal{K} = e^{i(\theta_{A} + \theta_{R})}\langle\Psi_{ee}|\Psi_{gg}\rangle$. The joint probability of measuring both detectors in the excited state is given by
\begin{align}
 P_{e_A e_R}(\tau) &= \text{Tr}[\rho^{\text{AR}}(\tau)|e_A e_R\rangle\langle e_A e_R|], \nonumber \\
 &= \frac{1}{4}\left[1 - V_{\text{joint}}\cos\left(\Omega_A \tau_A + \Omega_R \tau_R + \Delta\alpha\right)\right].
\end{align}
Here we defined the joint visibility as
\begin{align}
 V_{\text{joint}}(\tau) &= |\langle\Psi_{gg}|\Psi_{ee}\rangle|, \nonumber \\
 &= e^{-4\gamma_A(\tau_A) - 4\gamma_R(\tau_R) + 4 J_{AR}(\tau_A,\tau_R)}.
\end{align}
where 
\begin{align*}
    J_{\mathrm{AR}} = \lambda_A\lambda_R \text{Re}\int_0^{\tau_A}d\tau\int_0^{\tau_R}d\tau' G^{+}[x_A(\tau),x_R(\tau')],
\end{align*}
with $G^{+}(x,x') = \langle 0_M \vert\Phi(x)\Phi(x')\vert0_M\rangle$. In the asymptotic high-acceleration limit ($a\to\infty$), the rapid hyperbolic separation suppresses the vacuum-mediated spatial correlation term ($J_{AR} \to 0$). Consequently, the joint visibility reduces to the product of the purely local dephasing factors, 
\begin{align}
V_{\text{joint}}(\tau) = \e^{-4[\gamma_A(\tau_A)+\gamma_R(\tau_R)]}.
\end{align}

\begin{figure}[htp] 
    \centering 
    \includegraphics[width=0.45\textwidth]{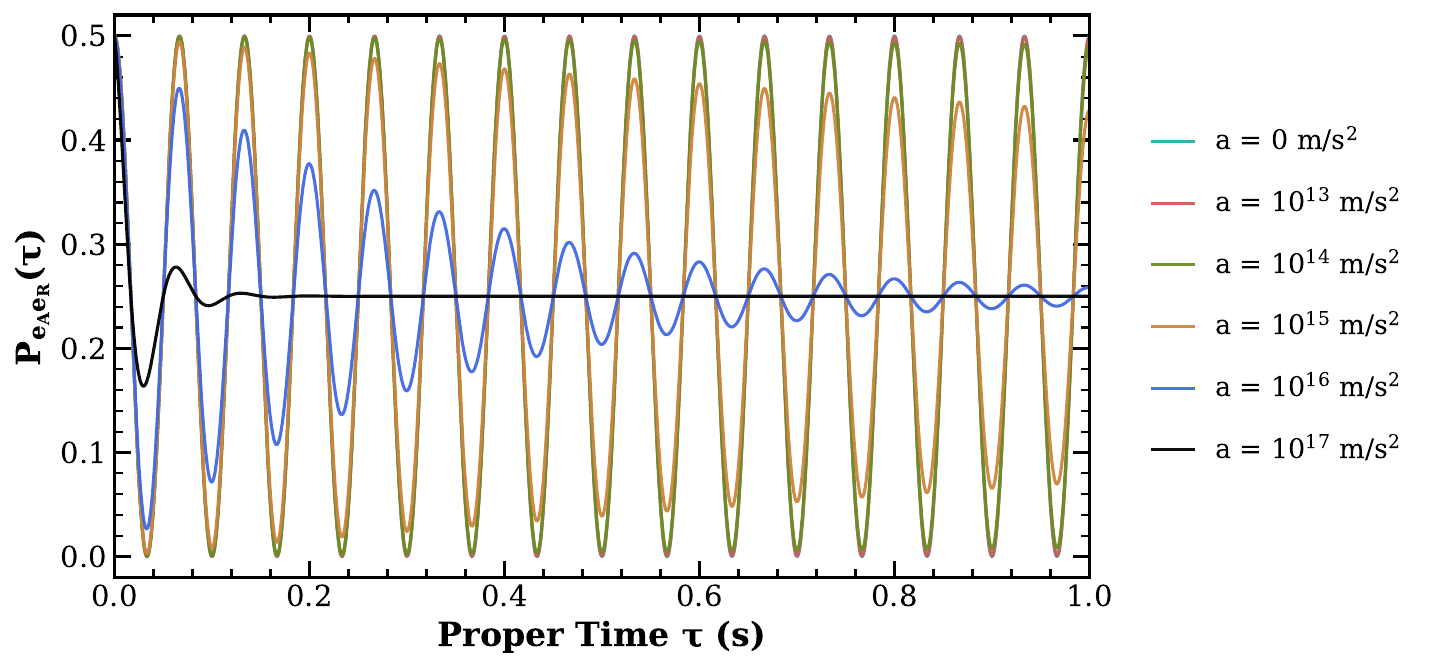}
    \caption{Joint probability $P_{e_A e_R}(\tau)$ as a function of the proper time $\tau$ for different values of Rob's proper acceleration $a$. The damping of the Ramsey fringes toward the classical statistical-mixture limit ($P=0.25$) reflects the progressive loss of coherence induced by Unruh thermal dephasing. Here, we adopt the parameters $\lambda=10^{-3}$, $\Delta\Omega=30\pi \text{ rad/s}$, and characteristic size $l \approx r_0 = 0.053 \text{ nm}$ (Bohr radius) in natural units ($c=1$).}\label{fig:probabilidade_conjunta_setup2} 
\end{figure}

 In Fig. \ref{fig:probabilidade_conjunta_setup2} we plot the joint probability $P_{e_A e_R}(\tau)$ as a function of proper time for different values of acceleration parameter $a$. When Rob is inertial ($a=0$), the probability exhibits smooth oscillations with fringe contrast remaining close to unity. As the proper acceleration increases ($a\geq10^{14}\,\mathrm{m/s^2}$), Unruh-induced thermal fluctuations enhance phase decoherence along Rob's branch, progressively reducing the amplitude of the joint oscillations. In the high-acceleration regime ($a\geq10^{16}\,\mathrm{m/s^2}$), the oscillatory pattern is rapidly suppressed, and the probability approaches the constant value, i.e., $P_{e_Ae_R}=0.25$, corresponding to the statistical limit of an incoherent classical mixture of four equiprobable states.

\begin{figure}[!ht] 
    \centering
    \includegraphics[width=0.45\textwidth]{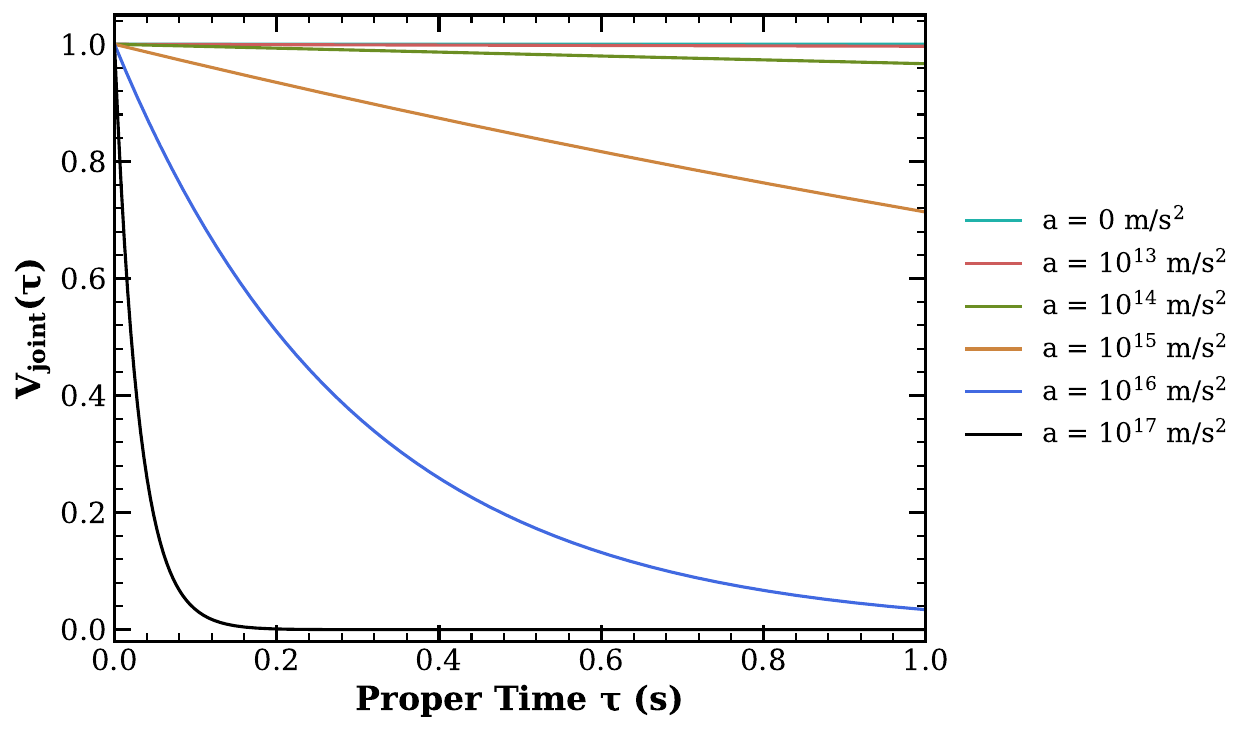}
    \caption{ Joint visibility $V_{\mathrm{joint}}(\tau)$ as a function of the proper time $\tau$ for different values of Rob's proper acceleration $a$. Here, we adopt the parameters $\lambda = 10^{-3}$, $J_{AR} \to 0$ (asymptotic spatial-decoupling limit), and characteristic size $l \approx r_0 = 0.053 \text{ nm}$ in natural units ($c=1$).}
    \label{fig:visibilidade_conjunta_setup2} 
\end{figure}

In Fig. \ref{fig:visibilidade_conjunta_setup2}, we plot the joint visibility $V_{\mathrm{joint}}(\tau)$ as a function of the proper time $\tau$ for different values of acceleration parameter $a$. It is shown that in the absence of acceleration ($a=0$), the loss of visibility is residual and governed solely by inertial fluctuations of the Minkowski vacuum. On the other hand, as Rob's proper acceleration increases, the thermal dephasing rate $\gamma_R(\tau_R)$ grows linearly with $a$, leading to a suppression of the joint interference contrast. 

Our goal now is to show that there is a connection between the joint visibility and the degree of entanglement between the two detectors. To do this, we will begin by calculating the unitary evolution of the system's state due to the interaction Eq. (\ref{HintAR}). After interaction, the initial state of the system $|\psi_1^{\text{AR}\phi}\rangle $ becomes 
\begin{align}
 |\psi^{\text{AR}\phi}(\tau)\rangle &= \frac{e^{-i(\nu_A + \nu_R)}}{\sqrt{2}}\Bigg[ e^{-i(\frac{\Omega_A\tau_A}{2} - \frac{\Omega_R\tau_R}{2} )}|e_A g_R\rangle\otimes|\Psi_{eg}\rangle \nonumber \\
 &+  e^{-i(-\frac{\Omega_A\tau_A}{2} + \frac{\Omega_R\tau_R}{2} )}| g_A e_R\rangle\otimes|\Psi_{ge}\rangle \Bigg],
\end{align}
where
\begin{align*}
 |\Psi_{eg}\rangle &= \prod_{\omega \textbf{k}_\perp} D_{\omega \textbf{k}_\perp}(\xi_{\omega \textbf{k}_\perp}^A - \xi_{\omega \textbf{k}_\perp}^R)|0_M\rangle, \\
 |\Psi_{ge}\rangle &= \prod_{\omega \textbf{k}_\perp} D_{\omega \textbf{k}_\perp}(-\xi_{-\omega \textbf{k}_\perp}^A + \xi_{\omega \textbf{k}_\perp}^R)|0_M\rangle.
\end{align*}
Again, by tracing out field degrees freedom $\rho_{\text{int}}^{\text{AR}}(\tau) = \text{Tr}_{\phi} \left[ |\psi^{\text{AR}\phi}\rangle\langle\psi^{\text{AR}\phi}|\right]$, the reduced density matrix of detectors can be written as
 \begin{align} 
    \rho_{\text{int}}^{\text{AR}}(\tau) = \frac{1}{2} \begin{pmatrix} 
    0 & 0 & 0 & 0 \\ 
    0 & 1 & \mathcal{K}_{AR} & 0 \\ 
    0 & \mathcal{K}_{AR}^* & 1 & 0 \\ 
    0 & 0 & 0 & 0 
    \end{pmatrix},
 \end{align}
with $\mathcal{K}_{AR} = e^{i(\Omega_A\tau_A - \Omega_R\tau_R )}\langle\Psi_{ge}|\Psi_{eg}\rangle$. Since the resulting matrix $\rho_{\text{int}}^{\text{AR}}(\tau)$ is an X-type state, it is well known that the degree entanglement can be quantified by Wootters concurrence, which is defined as 
\begin{align*} 
    C[\rho_{\text{int}}^{\text{AR}}(\tau)] = 2\max\left[0,\,\vert\rho_{23}(\tau)\vert-\sqrt{\rho_{11}(\tau)\rho_{44}(\tau)}\right]. 
\end{align*} 
Note that the pure-dephasing interaction preserves the populations of the outer diagonal states and the initial Bell state satisfies $\rho_{11}(0)=\rho_{44}(0)=0$, it follows that $\rho_{11}(\tau)=\rho_{44}(\tau)=0$ throughout the entire evolution. Consequently, the population-noise contribution vanishes identically, $\sqrt{\rho_{11}\rho_{44}}=0$, and the concurrence reduces directly to twice the magnitude of the coherence term, 
\begin{align*} 
    C[\rho_{\text{int}}^{\text{AR}}(\tau)] = 2\vert\rho_{23}(\tau)\vert = |\langle\Psi_{ge}|\Psi_{eg}\rangle|
\end{align*} 
Therefore, the loss of entanglement is directly determined by the distinguishability of the field states associated with the two branches of the superposition. In particular, at the spatial decoupling limit of the correlations between the detectors, when the cross-term contribution disappears. We obtain that the visibility of the two detectors fringes is the same as the concurrence of the state after interaction,
\begin{align}
  C[\rho_{\text{int}}^{\text{AR}}(\tau)] = V_{\text{joint}}(\tau), \quad J_{AR} \rightarrow 0. \label{Identiy}
\end{align}
This equality suggests that the interferometric contrast, $V_{\text{joint}}(\tau)$, provides a direct, operational, and quantitative measure of the bipartite entanglement between a pair of qubit. According to the identity Eq. (\ref{Identiy}), the collapse of joint visibility $V_{\mathrm{joint}}(\tau)$ directly tracks the decay of the bipartite entanglement induced by the effective Unruh thermal bath experienced by the uniformly accelerated detector.

\section{Conclusion}

In this work, we investigated acceleration-induced decoherence and entanglement degradation within a modified Ramsey-interferometric framework based on Unruh-DeWitt detectors coupled to a massless scalar field through a pure-dephasing interaction. The adopted coupling preserves the detector populations while allowing the quantum-field fluctuations to act directly on the phase coherence. For a uniformly accelerated detector of finite spatial extent, the resulting dynamics exhibits a clear separation between the residual decoherence associated with Minkowski-vacuum fluctuations and the additional contribution arising from uniform acceleration. In the long-time and high-acceleration regime, the latter becomes the dominant mechanism governing the suppression of coherence.

For the single-detector configuration, this loss of coherence is directly reflected in the damping of the Ramsey interference pattern. In the inertial regime, the fringe contrast remains essentially preserved for the parameters considered, whereas increasing the proper acceleration progressively suppresses the oscillations and drives the detection probability toward its incoherent limit. The Ramsey visibility therefore provides an operational probe of acceleration-dependent dephasing, translating the response of the detector to the quantum field into an experimentally accessible interferometric quantity.

The bipartite extension reveals a complementary effect on quantum correlations. When one detector remains inertial while the other undergoes uniform acceleration, the accelerated branch experiences an enhanced loss of phase coherence, which progressively reduces the contrast of the joint interference fringes. At sufficiently large accelerations, the increasing separation between the detector trajectories also suppresses the vacuum-mediated cross correlations, so that the dynamics becomes increasingly dominated by the local dephasing processes.

A particularly relevant result is the exact correspondence between the joint Ramsey visibility and the Wootters concurrence. Consequently, the degradation of the interferometric contrast directly tracks the loss of bipartite entanglement throughout the evolution. This establishes a transparent operational connection between relativistic decoherence and an experimentally accessible interference observable, allowing the surviving entanglement to be inferred from the Ramsey-fringe contrast within the present model.

Naturally, these results show that the proposed Ramsey-interferometric protocol provides a unified framework for characterizing both local coherence loss and acceleration-induced degradation of bipartite quantum correlations. Rather than focusing on the direct detection of field excitations, the approach encodes the effects associated with uniform acceleration in the loss of interferometric coherence. The analysis thus highlights Ramsey interferometry as a useful bridge between relativistic quantum-field phenomena and quantum-information observables.

\section*{Acknowledgment}

This work was supported by the Piauí Research Foundation (FAPEPI) and São Paulo Research Foundation (FAPESP) under Grant No. 002.144.FAPEPI.13072026 (EDCB-FAPEPI) and 2025/05176-7 (FCEL-FAPESP).

\section*{Conflicts of interest}

The authors declare no conflicts of interest.

\section*{Data availability}

No external datasets were used. The numerical data supporting the figures and table are generated from the equations presented in the manuscript.

\bibliography{refs}

\end{document}